%% file: main.tex
\documentclass[jornal]{IEEEtran}
\usepackage{xcolor, color, soul}
\usepackage{nicematrix}
\usepackage[numbers,sort&compress]{natbib}
\usepackage[utf8]{inputenc}
\usepackage[T1]{fontenc}
\usepackage[swedish,english]{babel}
\usepackage{graphicx, float, subfigure, blindtext}
\usepackage{hyperref}
\hypersetup{bookmarks=true,bookmarksnumbered=true, colorlinks=true,linkcolor={black},citecolor={black},urlcolor={black}}
\usepackage{mathtools}

\usepackage[nolist,nohyperlinks]{acronym}
\usepackage{cleveref}
\graphicspath{{images/}}

\usepackage{titlesec}
\titlespacing*{\section}{0pt}{*1}{*1}
\titlespacing*{\subsection}{0pt}{*1}{*1}
\renewcommand{\thesubsubsection}{\arabic{subsubsection}}
\titleformat{\subsubsection}[runin]{\itshape}{\thesubsubsection)}{1em}{}[:]
\titlespacing*{\subsubsection}{\parindent}{0pt}{*1}
 \usepackage {graphicx}

\author{\IEEEauthorblockN{ %
Moein Alavi,
Rasul Azizpour,
Hassan Zakeri, 
and,
Gholamreza Moradi $Senior Member$, IEEE
}

\IEEEauthorblockA{
Department of Electrical Engineering, Amirkabir University of
Technology \\
Email:
MoeinAlavi@aut.ac.ir, 
Ra.Azizpour@aut.ac.ir,
H.Zakeri@aut.ac.ir,
and, GhMoradi@aut.ac.ir
}}

\acrodef{svm}[SVM]{Support Vector Machine}

\title{Performance Improvement of True Time Delay Based Centralized
Beamforming Control with the Modulation Instability Phenomenon for Wireless Array Antennas}
\begin{document}
\maketitle
\input{sections/abstract}
\input{sections/keywords}
\input{sections/introduction}

\input{sections/Modulation_instability}
\input{sections/V_challenge_in_mach_zender}
\input{sections/True_time_delay_in_phase_array_antenna}
\input{sections/control_beamforming}
\input{sections/conclusion}

% Select the IEEEtran style
\bibliographystyle{IEEEtran}
%\bibliographystyle{apalike}
% Include bibliography file
\bibliography{IEEEabrv,refs}
\end{document}

%% file: sections/abstract.tex
\begin{abstract}

This paper proposes a novel method to enhance the performance of the modulator for fifth-generation wireless communication (5G) by exploiting modulation instability (MI). We show that MI can reduce the modulator’s bias voltage ($V_\pi$) by generating carrier side-band gain, and increase the modulation BandWidth (BW), resulting in higher channel capacity, without changing the modulator structure. In receive mode of the array antenna, where the signal is very weak, high-frequency amplification is a high demanding solution to mitigate coverage issue.

We also present a developed microwave-photonic beamforming bit-controller system for receiver-transmitter phased array antennas (PAAs), which are essential for high-capacity wireless communications like 5G. We employ a modulated frequency comb exploiting MI fiber to achieve an amplified true-time delay (TTD) technique for wide-coverage PAA beamforming and show that it can steer wideband high-frequency signal to a specific direction angle, avoiding beam-squint.

\end{abstract}

%% file: sections/keywords.tex
\begin{center}
\end{center}

%% file: sections/introduction.tex
\section{Introduction}
\label{section:intro}
Microwave Photonics (MWP) is the connection between opto-electronic and Radio-Frequency (RF) engineering. Photonic devices can generate and transmit radio-frequency signals with low loss, higher BandWidth (BW) and more flexibility for communication networks. Hence, MWP has become widely used in a variety of applications in recent years, including wireless systems like the Fifth-Generation (5G) mobile communication \cite{waterhouse2015realizing, romagnoli2018graphene}, optical telecommunications \cite{nureev2018microwave} medical imaging systems using THz waves \cite{zou2018microwave}, biomedical application \cite{mukherjee2019time}, and sensors  \cite{hervas2017microwave}. 
MWP systems convert RF signals to optical signals by Electrical-to-Optical (EO) modulators, transmitted through photonic devices like optical fibers and then extract RF signal from optical wave by photo-detectors. In MWP systems, modulation and photodetection are the most challenging parts, which degrade the system performance.

There are four fundamental performance challenges of the travelling-wave EO modulators:
I) microwave-optical velocity mismatch and excitation of undesired modes on the substrate, II) microwave electrodes lossess, III) high half-wave voltage ($V_\pi$) level, and IV) impedance mismatch of the electrodes and electrical system,
which leads to structure loss \cite{ren2019integrated,gopalakrishnan1994performance,presti2018intensity}. These challenges compromise the modulator's performance by decreasing the modulation response and BW.

Theoretically, the response of the EO modulation remains stable over the frequency of sidebands \cite{rueda2019resonant}. 
In the experimental case, essential factors like plasmonic losses, impedance mismatching, and microwave optical velocity difference, decrease the modulation response over frequency. Also, the product of modulation voltage and arm length in EO modulators is constant ($V_\pi L$=$\alpha$) in high efficiency mode. Consequently, the length of the EO arm increases by decreasing $V_\pi$.
According to the BW relation with the time constant, the response drops faster over frequency, and the BW decreases due to the EO arm. Therefore, the $V_\pi$ and BW of the modulators are directly proportional in typical cases. Therefore BW increasing leads to voltage increasing
\cite{gopalakrishnan1994performance,sun2018128}.

Decreasing microwave-optical velocity difference is one of the recent methods for increasing the BW of modulator systems while the $V_\pi$ remains constant \cite{rao2016high}.
In that case, $\alpha$ decreased by changing the structure of the modulator, type of excitation, and material of components. 
The effective light and microwave refractive index neared each other when the plasmonic dielectric and the time constant decreased, and BW rose. The complexity of a modulator increases, so the cost of the modulator structure and the BW range is not acceptable for new applications like 5G.

The authors of \cite{demirtzioglou2018frequency} used different resonators (such as ring resonators) and features of the EO circuit to create a structure with lower $V_\pi$. However, these structures are more complex and sensitive than others in device fabrication. Moreover, the resonant feature causes modulation response fluctuations in the sideband spectrum, which can distort the output signal.

The Fifth-Generation (5G) networks have been introduced to enhance the data rate and capacity of wireless communications. However, emerging applications such as autonomous driving, virtual/augmented reality, and the Internet of Things require higher capacity, which can be achieved by using high-frequency 5G. But, high-frequency signals suffer from high attenuation and path loss, which affect the signal quality and reduce the coverage area. Also, the modulator depth decreases over the sideband frequency, which makes the problem worse in photonic-based 5G systems. Therefore, to realize high-frequency 5G networks, low coverage issues should be addressed to ensure proper signal transmission and reception \cite{paul2022photonic,sudhamani2023survey}.

In this paper, we propose a novel method that uses the MI phenomenon in fiber optics to increase the performance of MWP-based T/R systems for 5G applications. Proposed method simultaneously improves the $V_{\pi}$ and BW of the modulator and amplifies RF signals at the high frequency sideband of the carrier. 

Beam-squint is a major problem in PAAs, which arises from the mismatch between the antenna's frequency and the PAA structure. To overcome this problem and enhance angular resolution in wideband applications, a TTD architecture has been proposed for high-capacity and low-latency wireless communication. We show that the modulation of the Frequency Combs (FC) is able to be enhanced by MI in fiber optics, which is applicable to MWP systems based on FC. We leverage the proposed MI structure to design a gained TTD system that eliminates beam-squint in wideband pulses. Therefore, the proposed method optimizes beamforming TTD structures based on MWP systems and facilitates high-frequency extensive coverage-range communication with high angular resolution.

Also, we propose a bit-controlled system that can control the directivity angle of PAA only with the length of the fiber.

%% file: sections/Modulation_instability.tex
\section{Fiber under Modulation instability}
\label{section:Modulation instability}
Modulation instability (MI) is a nonlinear phenomenon induced by the interaction of non-linearity, dispersion, and diffraction in the presence of an intense optical carrier wave. Rapid fluctuations appear as modulated pulse trains in most early articles, while the carrier is a continuous wave (CW) laser optical wave \cite{kudryashov2021model,agrawal2000nonlinear}.

The effects of fiber dispersion can be examined by using Taylor series to expand the propagation constant of the wave around the $\omega_0$:

\begin{equation}
    \beta(w)=\sum_{m=0}^{\infty}\frac{\beta_m}{m!}(w-w_0)^m,
    \label{taylorconstant}
\end{equation}
%The parameters $n(w)$ are group indexes, and $\beta_1$ and $\beta_2$ represent the group delay and dispersion of the group velocity, respectively.
where $\beta_m$ represents the $\mathrm{m}^{th}$ order of dispersion parameter. The second order ($\beta_2$) indicates the dispersion of the group velocity.

%The lowest-order non-linear effects in optical fibers are the third-order effects since they are constructed of silica, which has an amorphous microstructure. Nonlinear optical fibers exhibit various nonlinear phenomena and effects that influence the wave propagation in different spectrum channels. Among these phenomena, Self Phase Modulation (SPM) is a prominent one that reflects the nonlinear effect of a channel on itself, arising from the Kerr effect.

Optical fibers have third-order nonlinear effects from silica's amorphous microstructure. They exhibit nonlinear phenomena that influence wave propagation in different channels. Self Phase Modulation (SPM) is a prominent one that reflects the effect of a channel on itself from the Kerr effect.

The propagation of an optical pulse in optical fibers with sufficiently large lengths is influenced by both SPM and dispersion. The nonlinear Schrödinger equation:
\begin{equation}
\imath \frac{\partial A}{\partial z}+\imath \frac{\alpha }{2}A-\frac{\beta _2}{2}\frac{\partial^2A}{\partial T^2}+\gamma \left| A\right|^2A = 0
\label{eq:nonlinearshrodinger}
\end{equation}
where A, $\gamma$, and $\alpha$ are the slowly varying envelope of the optical pulse, the nonlinear parameter, and the loss factor of fiber, respectively. The above equation considers the effects of SPM and dispersion.

It can be shown that the envelope of a Continuous Wave (CW) laser is a soliton for the (\ref{eq:nonlinearshrodinger}). By neglecting losses for simplicity, the mentioned soliton is in the form of $\sqrt P_0 e^{i\gamma p_0 z }$, where $P_0$ and $\phi_{NL} = \gamma p_0 z$ are the incident power and the nonlinear phase shift induced by SPM, respectively. Stability of perturbed steady state CW leads to perturbations govern by:\begin{equation}
A = (\sqrt P_0 +a_1e^{i(Kz-\Omega t) }+a_2e^{-i(Kz-\Omega t) } )e^{i\gamma p_0 z }\end{equation}
form, where $K$ and $\Omega$ are the wave number and frequency of perturbation terms at sideband frequency of laser spectrum , respectively \cite{agrawal2000nonlinear}.
A non-trivial response necessitates that $K$ is related to the fiber parameters by:
\begin{equation}
K = \pm \frac{1}{2}\left|\beta_2 \Omega \right|\left [\Omega^2 +sgn(\beta_2)\Omega_c^2  \right ]^\frac{1}{2}, \; \Omega_c = 2/\sqrt{\left| \beta _2\right|L_{NL}} \end{equation}
which $L_{NL}=\frac{1}{\gamma P_0}$ is the nonlinear length of fiber. 
%($\Omega_c=\frac{2}{\sqrt{\left| \beta _2\right|L_{NL}}}$)
When $\beta_2$ > 0 (normal dispersion), $K$ becomes a real function of frequency. This means that the small perturbations are stable, but the $\beta_m$ values in (\ref{taylorconstant}) are altered and the spectral components have different propagation velocities. On the other hand, when $\beta_2$ < 0 (anomalous dispersion) and $\Omega < \Omega_c$, $K$ is imaginary and the perturbation amplitude grows along the fiber, which is called MI phenomenon. In this case, the $\beta_m$ values remain unchanged and the spectral components propagate in phase with the carrier wave. \color{black} 
The MI amplifies perturbation ($
G_{MI}(\Omega) = \left|\beta_2 \Omega \right| \sqrt{\Omega_c^2-\Omega^2}
\label{G_MI}
$).

\begin{figure}[!h]
	\centering\includegraphics[scale=0.59]{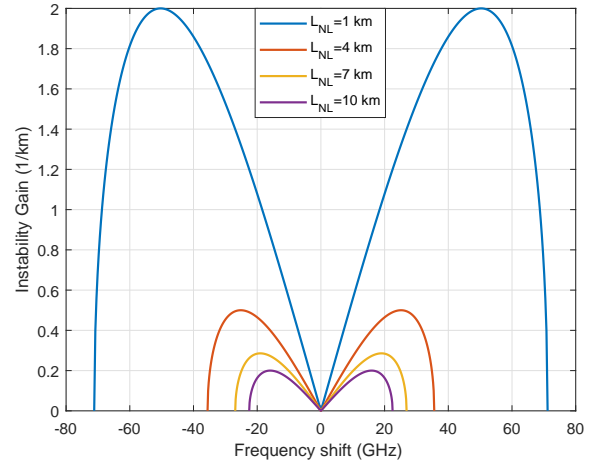}
	\caption{MI gain diagram for different nonlinear lengths of 1, 4, 7, 10 km. }
	\label{MI_gain}
\end{figure}

Figure \ref{MI_gain}  depicts the response of a 1km optical fiber  with $\beta_2 = -20 ps^2/km$ on the perturbation spectrum.

As shown in Fig. \ref{MI_gain}, a CW sideband can be amplified by a MI fiber within a confined high-frequency spectrum. The gain reaches its maximum at $\Omega_{m}\!\!=\!\!\frac{\Omega_c}{\sqrt{2}}\!\!=\!\!\sqrt{\frac{2\gamma P_0}{|\beta_2|}}$ frequency, which can be tuned to match different spectral ranges. The maximum gain of the MI fiber only depends on the ratio of the fiber length to the nonlinear length ($G_{MI}(\Omega_{m})\!\!=\!\! \frac{2 L}{L_{NL}}$). This amplifier also serves as a high-pass filter and does not require a separate pump wave \cite{devore2013enhancing}.
\begin{figure}[!h]
	\centering\includegraphics[scale=0.38]{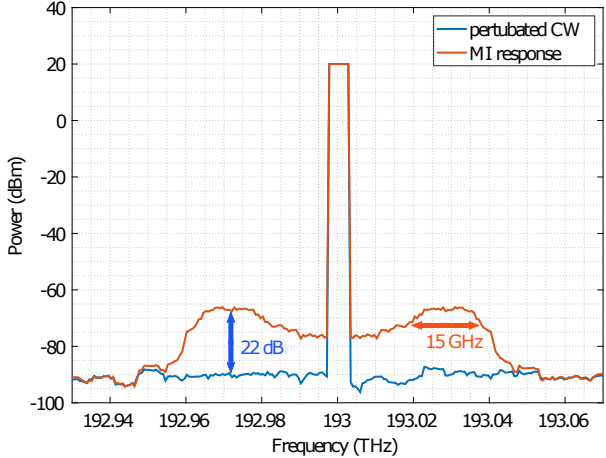}, \caption{Sideband response of a MI fiber optic with $\frac{L}{L_{NL}}$=3.23 km and $|\beta_2|\times L_{NL}\simeq$ 61 $ps^2$, to a CW perturbed by white noise. A narrow spectrum of the sideband with 15 GHz BW is amplified by 22 dB.}
	
	\label{compare_in_out}
\end{figure}

In high frequency long-distance wireless applications that use an array antenna, the signal level the receiver receives is lower than the signal level detected by the receiver. An improved gain performance can maintain the antenna's high transmission capabilities. When an antenna systems are worked for mm-waves 5G, the communication distances are drastically reduced due to the short wavelengths. 
Electromagnetic waves experience a higher signal quality and strength loss by atmospheric attenuation. In this case, achieving high gain is essential to reduce path loss in communication networks.
Therefore, the introduced structure can facilitate detection of the modulated signal by the receiver. 

Figure \ref{compare_in_out} shows the numerical response of a MI fiber to the CW laser with white noise.  According to MI relations, a 20 km fiber with $L_{NL} = 7.65$ km, increases the sideband level by about 22 dB. The amplification will continue as long as the sideband power satisfies the primary MI condition. If the sideband signal is too high, it will be distorted and the fiber will waste the laser power for unwanted nonlinear effects.\color{black}

\begin{figure}[!h]
	\centering\includegraphics[scale=0.38]{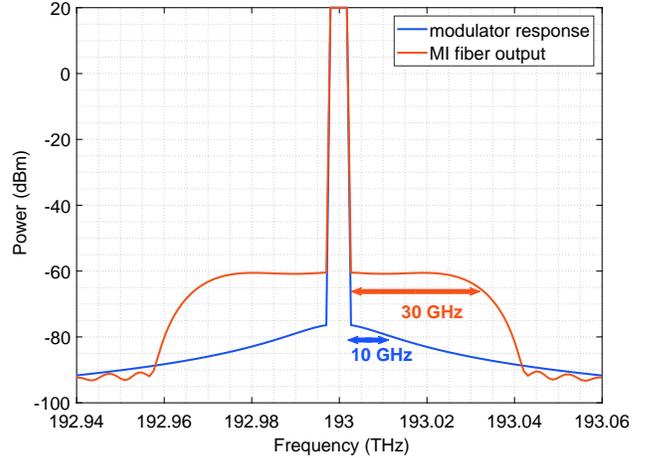}
	\caption{Bandwidth expansion of a modulator using MI fiber. A 22 km MI fiber, which is identified in Fig. \ref{compare_in_out}, compensates for the decrease of modulation depth and also increases the modulation depth. }
	\label{BWinc}
\end{figure}

As shown in Fig. \ref{MI_gain}, the MI-gain profile can offset the response degradation of conventional modulators over frequency. Figure \ref{BWinc} illustrates how MI can extend the BW of a 10 GHz MZM.

%% file: sections/V_challenge_in_mach_zender.tex
\section{ $V_{\pi}$ challenge in Mach Zehnder modulator}
\label{section:V_{pi} challenge in mach Zehnder}

To design a broadband travelling-wave Mach-Zehnder modulator (MZM), one of the fundamental performance challenges is the increasing of $V_{\pi}$ required for higher frequencies. In MZMs, $V_{\pi}$ is an index to determine the bias voltage needed for a acceptable modulation response. On the contrary, electronic devices that employ technologies such as CMOS exhibit a decrease in output voltage as the frequency increases \cite{golden2021reverse}.

\begin{figure}[!h]
	\centering\includegraphics[scale=0.3]{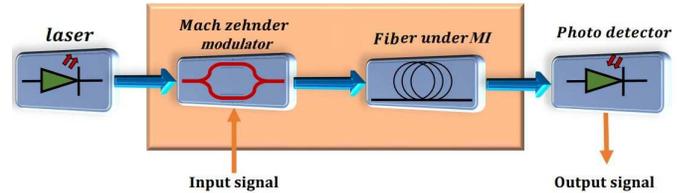}
	\caption{Block-diagram of a modulation part that exploited MI fiber.}
	\label{diagram}
\end{figure}

Figure \ref{diagram} shows the block diagram of the scheme that enhances the modulation performance by harnessing the MI fiber.

MI is stimulated by the low-power sidebands, which results in their expansion at the expense of the carrier. The RF output power is proportional to the $G_{MI}$, as the measured RF current is caused by carrier-sideband beating. 

In the amplified modulation of MZM, the proportional relation between bias voltage (V), input/output power, and MI gain is: 
\begin{equation}
P_{s,o}(w)\propto \frac{1}{V^2}P_{s,i}(w)G_{MI},
\label{Pout_V}
\end{equation}
where $P_{s,i}$ and $P_{s,o}$ are the input and output sideband power, respectively. Equation (\ref{Pout_V}) shows that necessary voltage for a certain modulation depth is decreased by using MI fiber. An increment of $P_{s,o}(w)$ due to $G_{MI}$ implies a corresponding decrement of $V_\pi$ by $G_{MI}^{1/2}$.

\begin{figure}[!h]
	\centering\includegraphics[scale=0.6]{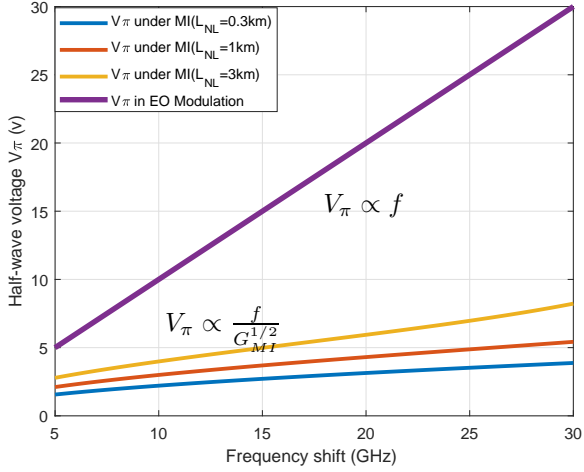}
		\begin{picture}(3,3)
\put(10,110){$V_\pi \propto f $}
\put(-50,65){$V_{\pi} \propto \frac{f}{{G_{MI}^{1/2}}}$}
\end{picture}
	\caption{Comparison of $V_\pi$ between MZM followed by MI fiber with different $L_{NL}$ and a simple MZM.}
	\label{half_wave voltage comparison}
\end{figure}

Figure \ref{half_wave voltage comparison} compares the $V_{\pi}$ required for a MZM gained by MI and also a simple MZM. By increasing the gain, bias voltage needed will be decreased. The MI phenomenon facilitates the integration of photonic and electronic devices, enabling the detection of weak high-frequency signals \cite{borlaug2014demonstration}. As shown, MI provides amplified sideband signal in higher BW  than simple MZM, without altering the modulator structure and design.

\color{black}

The noise figure (NF) of lightwave systems, which indicates the noise level and the transmission error probability, is linearly dependent on the $V^{2}_{\pi}$ of a MZM \cite{cox2006analog}. Therefore, reducing $V_{\pi}$ can lessen NF and prevent second-order distortion by setting the bias voltage to half of $V_{\pi}$. This enables the implementation of 5G systems with higher bandwidth, lower noise figure, and higher coverage range \cite{paul2022photonic}.

%% file: sections/True_time_delay_in_phase_array_antenna.tex
\section{True time delay in phase array antenna}
\label{section:True time delay in phase array antenna}

True Time Delay (TTD) is a crucial technique for addressing beam-steering and beamforming challenges in wideband PAA communication systems.
Optical devices can implement wide-BW TTD with less loss and electromagnetic immunity.

In 5G wireless communication, beamforming offers high array gains at both the transmitter (TxBF) and receiver (RxBF), resulting in increased SNR and more radio link margin that mitigates propagation route loss.
Multiple beams can be merged at the receiver to reduce path loss even more. Furthermore, interferences can be decreased, and frequency reuse can be improved using appropriate beamforming technology.

A single-fiber method \cite{xu2018photonic} uses a Frequency Comb (FC) source to achieve time delay between different channels in a dispersive fiber, rather than separate fibers that make structure bulky and complex. 
A fiber of length L transmits signals in distinct wavelengths $\Delta\lambda$ with time delay of
\begin{equation}
\Delta\tau=DL \Delta\lambda
\label{TTD_eq}
\end{equation}
where $D$ is the fiber dispersion. Pulses that modulated on FC carriers, experience time delay of $\Delta\tau=m T$, which $m$ and $T$ are difference of comb numbers and time delay of adjacent carriers, respectively. $T$ depends on Free Spectral Range (FSR) of FC. 

In the technique used by \cite{xu2018photonic} , PAA can steer pulse to
\begin{equation}
\theta_0=\sin^{-1}\frac{c.T}{d_{PAA}},
\label{angle_PAA}
\end{equation}

where $d_{PAA}$, $\theta_0$, and $c$ are spacing between radiated elements, steering angle and speed of light in the vacuum, respectively.

Equation (\ref{TTD_eq}) shows that the steering angle depends on two factors: the wavelength difference and the fiber length. The former factor can be changed by using a filter to select a frequency range from the FC. The latter factor can be changed by using different lengths of the same fiber type. This is more convenient and preferred.

\begin{figure}[!h]
	\centering\includegraphics[scale=0.36]{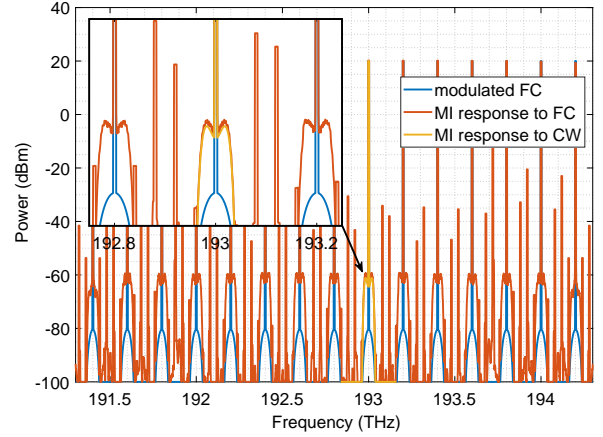}

	\caption{Performance of the MI fiber on a modulated frequency comb. The MI fiber amplifies the sideband of FC carriers, same as a modulated CW in 193 THz. FWM generates unwanted components in the out of MI band.}
	\label{fig:comb1}
\end{figure}

To account for the XPM effect in a multi-channel setup \cite{agrawal2012fiber}, the nonlinear Schrödinger equation has to be modified \cite{agrawal2000nonlinear} by replacing $\gamma {|A_j|}^2 A_j$ with $\gamma ({|A_j|}^2+2\sum_{k\neq j} {|A_k|}^2 ) A_j$ in (\ref{eq:nonlinearshrodinger}), where $A_m$ is the signal of the $m^{\text{th}}$ channel. The response of a nonlinear dispersive fiber to uniform FC is considered, and the variation of $\beta_2$ in the wavelength domain can be ignorable. The equal power of carriers in shaped FC leads to same phase variation of all steady-state solitons.
Using the modified equation for the carrier perturbation of the $j^{\text{th}}$ channel in the FC, we obtain $a_j\!\!=\!\!(u_j e^{i(Kz-\Omega t)}+ i v_j e^{-i(Kz-\Omega t)})e^{i\Phi_{\mathrm{XPM}} z}$, where $\Phi_{\mathrm{XPM}}$ is the phase variation caused by XPM. The equations have a non-trivial solution when $K^2=\omega^2 \frac{\beta_2}{2} (\omega^2 \frac{\beta_2}{2} + 2 \gamma P_0)$ for all FC channels. Therefore, similar to a single CW, a perturbed FC can stimulate MI phenomenon in an anomalous dispersion fiber. The XPM effect causes a uniform increase in the phase accumulation of all combs.

Another nonlinear effect that we do not want in a multi-channel setup is the Four-Wave Mixing (FWM) effect \cite{agrawal2012fiber}. We can ignore the FWM effects in fibers with more than 5$\frac{\text{ps}^2}{\text{nm}}$ of dispersion. Also, the MI condition on the sideband power limits the power of the FWM components. Moreover, if the MI-response spectrum covers only half of the frequency comb spectrum, we can avoid producing significant FWM components in the desired band \cite{agrawal2012fiber}. \color{black}

The most unwanted nonlinear effect that limits the performance of the parametric amplification in a nonlinear fiber is the Stimulated Brillouin Scattering \cite{coles2010bandwidth}. This effect reduces the power of the carrier in the fiber and produces undesirable reflected components. It can be suppressed by keeping the laser power below a certain threshold. One way to increase the threshold of Brillouin Scattering is to apply phase modulation. This can raise the threshold up to 100 mW \cite{borlaug2014demonstration}, which is the value we use in our parameter settings.\color{black}

To benefit from MI-gain, the parameters have to be set so that the signal spectrum aligns with the MI-gain band. This prevents the noises outside the band from being amplified and preserves the Signal-to-Noise Ratio (SNR) after the amplification. If the signal spectrum is outside this band, the MI boosts the noise more than the signal and reduces the system's sensitivity. \color{black}

The nontrivial solution of the modified Schrodinger equation cannot be split if the carriers in a FC have different powers. This makes the response of each carrier depend on the power of the others and causes unequal amplification of the signal. This reduces the amplification available on the signal and increases the noise, which can worsen the performance. Moreover, the power imbalance changes the phase variation of channels of each carrier. However, this does not affect the wave response of the channels, which follow their corresponding carrier. A wave-shaper can be used to adjust or equalize the power distribution among the carriers in the FC \cite{wu2010generation}.\color{black}

Figure \ref{fig:comb1} displays the response of the anomalous nonlinear fiber, shown in Fig. \ref{compare_in_out}, to a modulated single FC. It reveals that the MI is stimulated by the perturbed FC carriers. The MI fiber boosts a narrow high-frequency spectrum of the sidebands, which peak at frequencies that are inversely proportional to $\sqrt{\beta_2}$.
The weak and negligible components that appear outside the MI-band are due to the FWM effect. In a frequency comb-based TTD, a filter eliminates these unwanted components and extracts the pulses delayed by the system.\color{black}

These indicate that the improvements come from MI in the single CW are able to extend to systems that operate by FC source. This opens up new possibilities in MWP based applications that require high-performance modulation. For instance, it can mitigate the high propagation-loss and low MZM response of high frequency signals in wireless communications, as well as its potential to increase BW in analog links \cite{paul2022photonic}.

\color{black}

%% file: sections/control_beamforming.tex
\section{beamforming bit-controller system}
\label{section:results}

Multiplexing techniques in the time, frequency, code and spatial domains have been essential research areas to meet the increasing demand for spectral efficiency in 5G mobile wireless communication systems.
With a sufficient correlation between the PAA, adaptive beamforming applies to closely spaced antenna arrays. Therefore it is possible to improve network coverage, increase signal quality and exploit the array gain.

\begin{figure}[!h]
\centering\includegraphics[width=10cm,height=5.5cm]{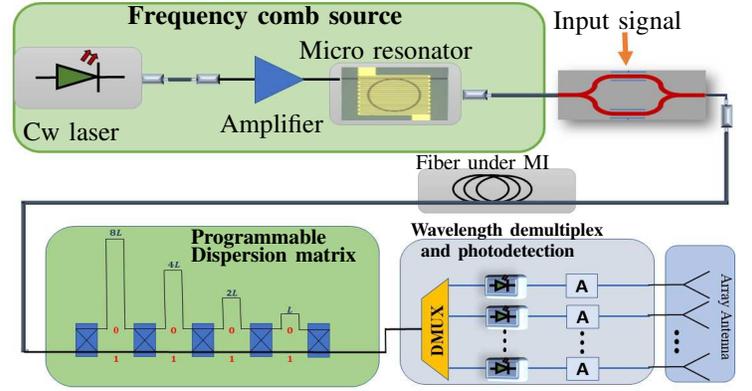}
\begin{picture}(3,3)
\put(-118,113){$\text{ Cw laser} $}
\put(-40,116){$ \text{Amplifier}$}
\put(-10,144){$ \text{Micro resonator} $}
\put(-75,157){$ \textbf{Frequency comb source} $}
\put(86,155){$ \text{Input signal} $}
\put(34,102){$  \text{\footnotesize Fiber under MI} $}
\put(-51,74){$ \textbf{\footnotesize Programmable
} $}
\put(-51,67){$ \textbf{\footnotesize  Dispersion matrix} $}
\put(+32,77){$ \textbf{\scriptsize Wavelength demultiplex} $}
\put(+36,69){$ \textbf{\scriptsize  and
photodetection} $}
\end{picture}
	\caption{Structure of modulation instability with micro comb based microwave TTD beamformer with the bit-controller-based length of the fiber.}
	\label{microcomb structure}
\end{figure}

Figure \ref{microcomb structure} demonstrates a microcomb-based TTD beamformer with bit-control. Pumping a non-linear microresonator with a continuous wave laser generates a broadband optical FC. A MZM modulates the microwave signal on FC, which is then followed by an MI fiber.  The flatness of the FC is a crucial factor for obtaining the optimal performance improvement in photonic-based 5G systems. A wave shaper can adjust the power distribution among the carriers in the FC to make it more flat. If the FC is less flat, the available amplification on the signal decreases, while noises are amplified undesirably. Moreover, this can alter the phase variation of channels of each carrier. However, this issue does not significantly affect the dispersion order due to the wave response of the MI, and therefore, it can still be used for MWP systems based on FC. Nevertheless, the overall performance improvement is reduced. \color{black}
\begin{figure}[!h]
	\centering\includegraphics[scale=0.4]{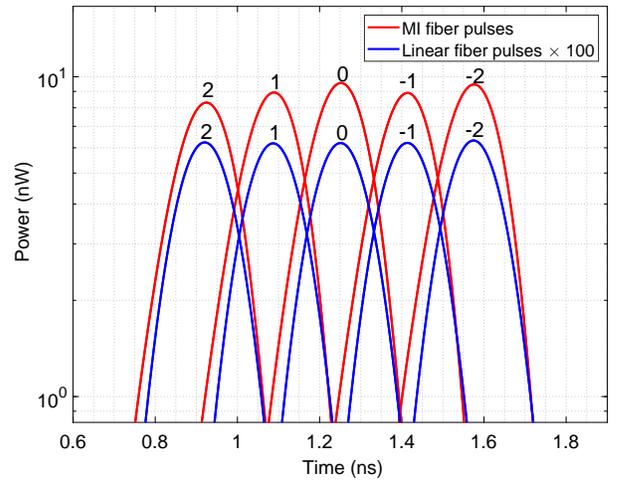}

	\caption{Temporal response of a FC modulated by a 4 GHz pulse in a MI optical fiber and a linear optical fiber. The numbers above the pulse peaks indicate the corresponding comb number $n:(f(\text{THz})=193+n*\text{FSR})$, where FSR=200 GHz. The MI fiber produces pulses with constant delays and also enhances the modulation depth by more than 100 times.}
	\label{fig:Timedelays}
\end{figure}

As we discussed before, the MI phenomenon preserves the wave properties of the sideband components during propagation, just like a linear fiber. Figure \ref{fig:Timedelays} displays the temporal response (red) of a 0.8 km MI optical fiber, when a 4 GHz pulse at a frequency of 28 GHz is modulated on the carriers. For this analysis, we used a highly nonlinear anomalous fiber with $\gamma=36 \frac{1}{\text{W.km}}$ and $|D|=127\frac{\text{ps}}{\text{nm.km}}$ at 1550 nm.  The highly nonlinear photonic crystal fibers can achieve the desired dispersion amount in a certain operating wavelength by adjusting the zero-dispersion wavelength \cite{knight2000anomalous}. Liquid-core fibers are also suitable for this application, as they offer higher nonlinearity and tunable $\beta_2$ \cite{chemnitz2023liquid}. Fibers with a higher nonlinear coefficient can decrease the power needed for carriers and avoid using an optical amplifier. Optical amplifiers introduce unwanted noises and reduce the system's sensitivity.   

The figure demonstrates that the MI fiber produces pulses with similar delays as the linear fiber response (blue), while also enhancing the modulation depth by its unique advantages. The MI fiber, with its low SNR degradation \cite{borlaug2014demonstration}, is a suitable option for improving the sensitivity of MWP-based high-frequency 5G networks, especially in array antenna architectures. This can lower the number of stations needed and the network latency.\color{black}

We use the normal fiber with programmable length to control the time delay for PAA by tuning the minimum time steps. The programmable part can provide discrete lengths from $L$ to $15\times L$. By setting the smallest piece of the normal fiber appropriately, we can align the beam steering angle with the zero reference.

\begin{figure}[!h]
	\centering\includegraphics[scale=0.6]{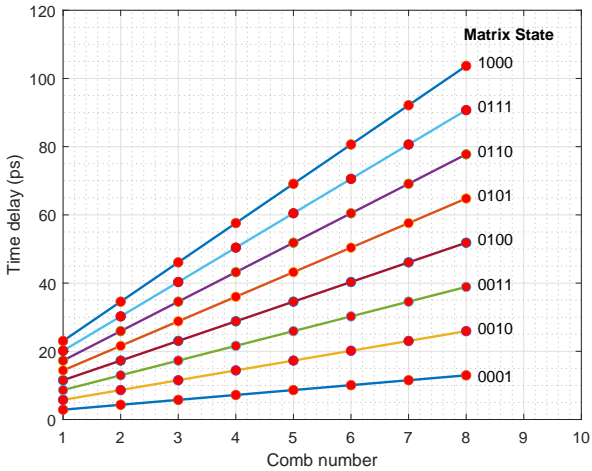}
	\caption{Time delays of the comb lines with the dispersion matrix through the states from “0001” to “1000”.}
	\label{fig:combdelay}
\end{figure}

The dispersion matrix is made of a binary delay line of optical switch devices and dispersive fibers (Fig.\ref{microcomb structure} shows an example with four bits). Then, each carrier's channel is photo-detected and de-multiplexed. Fig.\ref{fig:combdelay} shows the time delay state for PAA with the minimum time delay $\Delta\tau = 1.44 ps $ (This is the minimum time delay created by the minimum fiber length). We can change the binary code to tune and control the time delay and steer the PAA main beam.

\begin{figure}[t]
\centering
\subfigure[M=4]{\label{fig:figure:im1} %% label for first subfigure
\centering\includegraphics[clip,trim={4cm 3.5cm 2.5cm 3cm},scale=0.3]{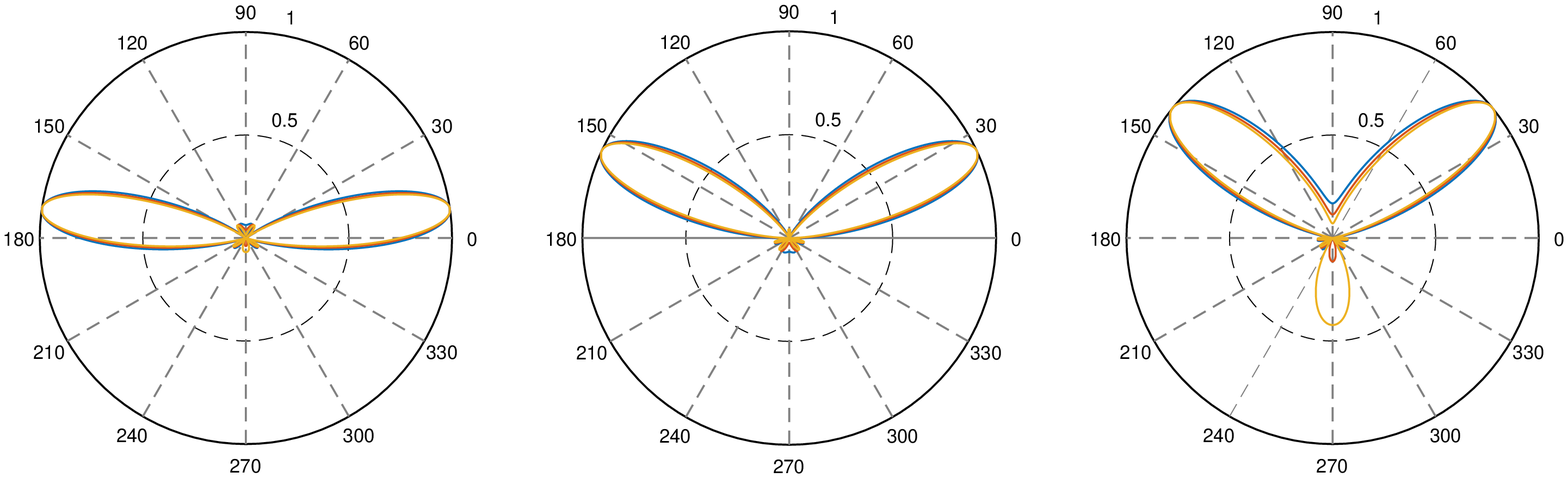}
}
\hspace{6mm}
\subfigure[M=8]{\label{fig:figure:im2} %% label for second subfigure
\centering\includegraphics[clip,trim={3.8cm 8cm 2.5cm 7cm},scale=0.32]{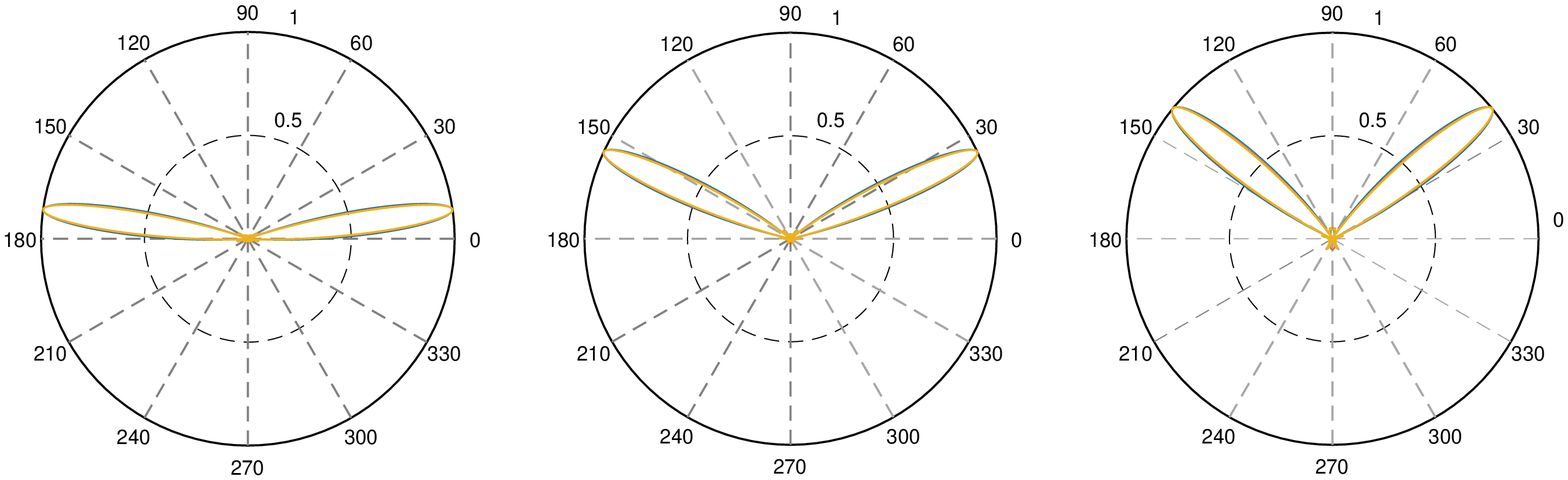}
}
\caption{Array patterns for 26, 28, and 31 GHz frequencies in three bit-settings that steer the beam to 8°, 24°, and 40° angles, using a) four and b) eight elements. The figure shows that the proposed structure can beamform different frequency components to the same direction, without beam squint.}
\label{fig:figure} %% label for entire figure
\end{figure}

Based on the 4-bit model, the target-controlled fiber in the dispersion matrix is 79 m, 39.5 m, 19.75 m, and 9.875 m.
Fig.\ref{fig:figure:im1} and Fig.\ref{fig:figure:im2} shows the simulated radiation patterns for 4 and 8 array antennas for three different cases at 26, 28, and 31 GHz, respectively.

\color{black}

The equation (\ref{angle_PAA}) determines the beam steering angle of the TTD architecture. The array pattern ($G_a(\theta,\lambda)$) of the PAA elements that receive uniform delayed pulses is calculated as follows \cite{longbrake2012true}:
\begin{equation}
G_a(\theta,\lambda)=\frac{\sin^2[M\pi(d_{PAA}/\lambda )(\sin\theta -\sin\theta_0)]}{M^2 \sin^2[\pi(d_{PAA}/\lambda )(\sin\theta -\sin\theta_0)]}.
\label{pattern_eq}
\end{equation}

Figure \ref{fig:figure} shows array factor of the bits controlled to steer the direction of the 28 GHz component to 8$^{\circ}$, 24$^{\circ}$, and 40$^{\circ}$ is shown for different number of elements. The antenna's phase offset was tuned to steer the peak lobe at a specific angle. The $d_{PAA}$ is $\frac{\lambda_{rf}}{2}=5.4 mm$. The  $\omega_0$ is the angular reference frequency, which is 2$\pi\times$193 THz. The array antenna pattern becomes narrower and the back lobe effect decreases as the number of radiating elements increases. Therefore, for high-technology wireless communication, it is better to increase the number of antennas to achieve accurate results.

The MI-based MWP allows the proposed TTD architecture to beamform the low-voltage, wide-BW signal at high-frequency without beam squint.

%% file: sections/conclusion.tex
\section{Conclusion}

In this paper, we have presented and verified a novel modulation system using MI for 5G applications. We have demonstrated that MI can lower the bias voltage needed for modulating high-frequency components by amplifying the sideband frequency. The MI provides a 22 dB gain with an appropriate BW for 5G networks and also considerably expands the BW of a 10 GHz modulator. The MI can enhance the channel capacity for wireless communication applications.

We have further shown that the MI phenomenon can be utilized for the modulated FC. Due to the wave-response preservation of MI, the MI fiber can implement a FC-based MWP system with improved performance. We have devised a TTD architecture using a modulated FC amplified by MI, which also resolves the beam-squint issue. The beam was steered with a bit-control system that altered the fiber length. We have realized three array patterns in 8$^{\circ}$, 24$^{\circ}$, and 40$^{\circ}$ directions by adjusting the bit-control system to suit the 26, 28, and 31 GHz components.

This work illustrates the potential of enhancing the performance of MWP-based 5G network beamforming by exploiting MI phenomenon. Future research can investigate other applications and challenges of MI fiber in wireless communication and also examine the MI in smaller platforms, such as integrated photonics, to address the size issues.